\documentclass{ws-jai}
\usepackage{enumitem}

\newlist{counting}{enumerate}{10}
\setlist[counting]{
	label*=\arabic*.,
  	labelindent=1\parindent,
  	itemindent=0pt,
  	leftmargin=1\marginparwidth,
	rightmargin=\dimexpr\linewidth-14cm-\leftmargin\relax,
	before=\setlength{\listparindent}{-\trimwidth}
}

\begin{document}

\title{The Swarm Telescope Concept}

\author{Jayce Dowell$^{1,2}$ and Gregory B. Taylor$^1$}

\address{$^1$Department of Physics and Astronomy, University of New Mexico, 
Albuquerque, NM 87131, USA\\
$^2$ jdowell@unm.edu\\
}

\maketitle

\abstract{
As telescope facilities become increasingly more capable they also become increasingly complex and require additional resources to operate.  This is particularly true for the current and future generations of ``software defined telescopes" that can support a variety of observing programs simultaneously, either through commensal observations or through support for multiple pointing centers as in the case of dipole arrays or dishes equipped with phased array feeds.  At the same time, many current and future facilities are also distributed over large geographic areas, making monitoring and maintenance more difficult and costly.  For these reasons we have developed a new paradigm for telescope operations called the ``swarm telescope" that breaks large, single facilities into smaller groups of independent systems that can collaboratively work together to function as a single facility but with much less operational overhead.  In this paper we outline the swarm telescope concept and an example of its implementation at the Long Wavelength Array.  We also discuss potential advantages of using this approach for other facilities, in particular the Next Generation Very Large Array.
}


\section{Introduction}
Traditionally, astronomical observatories across the electromagnetic spectrum have focused on the construction, operation, and maintenance of a single telescope.  Even in cases where an observatory operates multiple telescopes, such as one of the arrays of the national observatories, the facilities are usually clustered in a small geographic area and operated through independent headquarters, such as the case for the National Optical Astronomy Observatory Kitt Peak Cerro Tololo Inter-American Observatories.  However, in radio astronomy there are several examples of facilities that are spread over an extensive geographic area in order to obtain high angular resolution.  This dispersal adds to the construction and maintenance costs associated with these facilities and makes them more difficult to operate due to differing observing conditions at each site which may preclude higher frequency observations using the entire array.  The Very Long Baseline Array \citep[VLBA;][]{vlba} as well as the proposed Square Kilometer Array \cite{ska} are two such examples of facilities that are dispersed over numerous sites covering a large geographic area.

Radio observatories are also facing a separate but equally challenging problem through recent advances in computing and digital signal processing.  It is becoming increasingly common for a telescope to carry out multiple simultaneous observations with different goals, either through the use of commensal observing or through multiple independent pointing centers.  This second approach is especially common in low frequency dipole arrays where each element of the array is sensitive to a large fraction of the visible sky and multiple beams can be formed by coherently adding these signals in different ways.  Similarly, dishes equipped with phases array feeds, e.g., \citet{paf1,paf2}, are also also capable of observing multiple targets at once although the field of view is typically more limited by the gain pattern of the dish.  Thus, one telescope can become two or more.  Although this is a boon for transient searches and other experiments that require large amounts of observing time, it increases the operational complexity of the facility due to conflicting configuration requirements and scheduling considerations.

Motivated by these challenges to facility operations and looking forward to the next generation of radio facilities, we have developed the swarm telescope concept for facility operation.  This concept is an attempt to balance the desire to use all of the new technological capabilities for observing multiple independent observation simultaneously without substantially increasing the operational and maintenance overhead for distributed arrays.  At the core of the swarm telescope is a shift away from thinking about an observatory as a monolithic entity.  Rather, an observatory is viewed as many independent parts that work together to accomplish scientific observations.  This shift requires moving part of the decision making about the facility away from the human schedulers and operators and transitioning it to ``software defined operators" that run on each part of the facility.  These software agents then communicate with each other and build dynamic arrays to accomplish the goals of multiple observers, while also adjusting for varying observing conditions and array element states across the facility.  This is inspired by the concept of swarm intelligence proposed by \citet{SI} where independent agents interact with each other and their environment to perform complex behaviors.


The idea of automation and dynamic scheduling is not new in astronomy and efforts to increase the efficiency of facilities range from queue observing, e.g., the Gemini Observatories \citet{GeminiQueue}, up to fully robotic telescopes that conduct observations with little human intervention,  such as the telescopes of the e-Science Telescopes for Astronomical Research project \citep[eSTAR;][]{eStar} but also see \citet{baruch1992, CollabTelescope} for further examples.  Building on the concept of robotic telescopes are so called ``thinking" telescopes, e.g., Rapid Telescopes for Optical Response \citep[RAPTOR;][]{raptor} which use machine learning to not only operate the facility but to also analyze the data and make decisions about future observations.  The swarm concept expands upon the building blocks of queue observing and robotic facilities and combines them with the advantages of interferometry.  This allows for the operation of a dynamically scheduled facility, with greater resolution and sensitivity than the individual components, that also has flexible resource allocation to maximize observing time and reduce manual operations and maintenance requirements.  Although the development of this approach was motivated by future large facilities, it also has implications for smaller, university-lead projects.

In this paper we outline the general components of the swarm telescope concept in \S\ref{sec:core} and provide a concrete example of the concept as it is implemented at the Long Wavelength Array (LWA) in \S\ref{sec:implementation}.  Section \ref{sec:examples} provides an example of how this concept may be applied to the specific case of the Next Generation Very Large Array while \S\ref{sec:conclusion} provides a summary of the concepts.

\section{Core Components}
\label{sec:core}
The swarm telescope concept is designed with the primary goal of reducing the amount of human interaction required to operate a radio interferometer.  Under the swarm concept the interferometer is managed not as a single entity but as a collection of independent and autonomous elements that are connected together though a data transport system.  For this type of swarm facility there are three fundamental components:  autonomous element control, a method of inter-element communication, and data transport management.  The role of each of these is discussed in detail below.


\subsection{Autonomous Element Control}
The most critical component of the swarm concept is autonomous element control which governs the actions of each element of the facility.  The function of the autonomous element control may more appropriately be thought of as a ``software defined operator"  that must respond automatically to changing conditions in addition to carrying out pre-defined observations.  This is similar to what is used to control a single robotic telescope and differs from a traditional monitor and control systems that only implements environmental monitoring, equipment/instrument monitoring, and logging.  


The primary function of the element control is to ensure the safety of the telescope and its instrumentation while also maximizing the utilization of the element as part of the interferometer.  This principle drives the requirements of the control system in two different directions.  The first, safety of the element, requires multiple monitoring points and preventative actions in order to identify and prevent problems.   The second direction requires methods of relating the goals of an observation to the performance of an element in order to maximize the quantity and quality of the observations, and automated methods of recovering from problems when they occur.  An ideal control system would therefore:

\begin{counting}
	\item Have a variety of monitoring points, both internal and external, that can be used to determine the operational state of the element using a combination of the current and historical sensors values.  These would then be combined to create a simple ``go, no-go" determination of the overall operational state of the element.
	\item Use the current go, no-go state in order to implement shutdown and recovery procedures as appropriate.  For example, shutting down the element to protect the equipment from damage due to an approaching thunderstorm or recovery of the element after a power outage.  Where automated recovery is inappropriate, such as in the case of a mechanical failure, the control system should send a notification to the observatory that manual intervention is required, while putting the element in a safe stowing state as necessary.
	\item Use the monitoring points and go, no-go state combined with the observing schedule to determine which observations to execute with the goal of optimizing the available resources and maximizing observing time.  This can also be expanded to on-the-fly trigger processing where scheduled observations can be interrupted to observe a target of opportunity.  For elements that can carry out multiple primary observations the control systems should be able to independently monitor and balance observations across the available resources.
	\item Be robust against any errors that may occur while performing any of the actions listed above, either within the element control software or as part of a hardware failsafe system.
\end{counting}

The details of the control system implementation is largely dependent of the nature of the telescope and the desired level of complexity for its control.  For example, dipole arrays with no moving elements will require different monitoring points and procedures than steerable dishes that have travel limits and are subject to environmental factors such as wind loading.  Similarly, the time required for automated shutdown and recovery is dependent on whether or not the elements are movable, how many systems need to be shutdown or put into a low power state, and amount of time needed for self-diagnostics once the shutdown condition has passed.  These consideration may also require minimum downtimes in order to keep systems from being rapidly shutdown and restarted.  Finally, another critical consideration in the structure of the control system is the safety of any people who are on site.  This adds an additional layer of complexity to the control structure since it requires interfacing with hardware lockouts in order to prevent any unintended or unsafe operation.

These implementation considerations also govern the level of intelligence designed into the control system.  For beamformed dipole arrays it may be sufficient to use a set of heuristics and state machines to determine the operational status of the element.  A specific example of how control is implemented for a dipole array is presented in \S\ref{sec:hal} for the Long Wavelength Array.  Steerable dishes, on the other hand, may benefit from more advanced artificial intelligence and machine learning techniques in order to account for any mechanical hysteresis in the system.  It should also be noted that more advanced machine learning methods may also lead to greater reductions in operations and maintenance costs by being able to optimize each element independently.  This can range from methods of predicting component failure before it impacts the element to accurate local weather forecasting.  However, it should be noted that in some situations artificial intelligence and machine learning techniques may not be desirable since they add to the complexity of the control and can result in unexpected and non-deterministic behavior of the control system.  This is particularly true for evolutionary algorithms, e.g., see \citet{aiwoes} and references therein, where the exact nature of the fitness metric may reveal edge cases that do not produce the intended function.  An understanding of the extent of non-deterministic behavior and how to limit it is essential.  We also note that the many different approaches that can be implemented within the control system comes with an associated development and testing cost.  From a facility standpoint this can be thought of as effectively shifting long-term operating costs into the initial construction costs.


\subsection{Inter-Element Communication}
\begin{figure}
	\centering
	\includegraphics[width=3.3in]{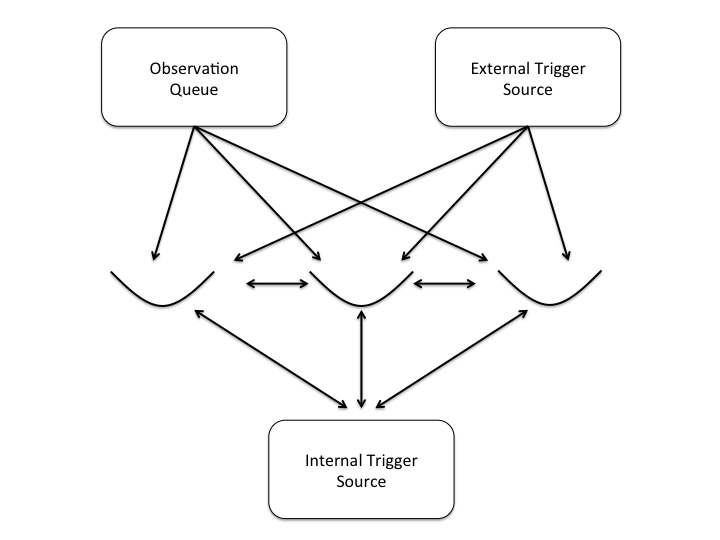}\hspace{1ex}
    \includegraphics[width=3.3in]{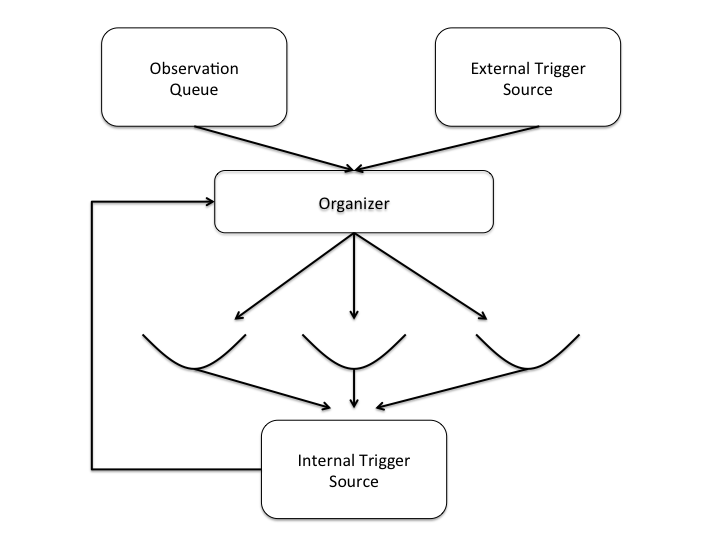}
	\caption{\label{fig:communication}Comparison between the two types of inter-element communication in the swarm concept.  The diagram on the left shows a leaderless system while the right diagram shows an organizer system.  As the name implies, the leaderless system has no single point of control.  All of the communication is done through broadcast messages, either from an external source of observations or event triggers such as gamma ray bursts, or through self-generated triggers from real-time transient searches.  In the organizer system all of the communication between the elements and with the observation queue is done through a single point of control, the organizer, so that a global picture of the array state can be maintained.}
\end{figure}

Inter-element communication is what allows the individual elements of an array to coordinate and form the interferometer.  Figure \ref{fig:communication} shows the two general types of inter-element communication that can be implemented in the swarm telescope framework:  leaderless and organized.  In a leaderless system there is no single point of control for the combined array.  Rather, the elements communicate directly with each other or external agents, and conduct observations given the operational state and available resources.  This system represents the most minimal form of communication and is best suited for arrays with small numbers of elements and for observing targets of opportunity such as gamma ray bursts, gravitational wave triggers, or self-generated trigger events from transient searches using array elements that can support multiple simultaneous observations.  The LWA uses the leaderless approach, which is discussed in \S\ref{sec:lwaops}.

For larger arrays or arrays where the elements cannot support multiple simultaneous observations, an organizer framework allows more flexibility for the synthesized interferometer.  In this approach there are one or more organizer agents added to the inter-element communication system.  These organizers poll the individual elements and maintain a global state for the array.  These organizers can then combine this information with the contents of the observing queue in order to choose specific observations and build on-the-fly arrays that carry out those observations given the constraints of varying weather conditions across the array, inoperable elements, required sensitivity, etc.  This polling and triggering process allows for adjustments to be made to the observing program to reflect the current state of the array.  For example, the duration of each observation can be adjusted to obtain the required sensitivity given the number of elements that are on-line.  The organized framework also allows for other multi-element behaviors to be implemented, such as phasing up nearby elements to increase sensitivity on certain baselines or breaking the array into multiple sub-arrays.  Some of these behaviors are discussed in more detail in \S\ref{sec:examples} in the context of the Next Generation Very Large Array.  

Although the swarm concept does allow for a greater degree of flexibility for scheduling observations it does come at the cost of increased observation accounting.  This is most obvious in the situation where observations are interrupted by a trigger, and an observation is halted before the goals of the observation are met.  In order to deal with these situations, information about the state of the interrupted observations, such as the current estimated sensitivity and accumulated $(u,v)$ coverage, needs to be incorporated back into the observation queue so that these observations may be completed at a later time.  Thus, the observing queue used by the organizers would contain not only information about the target and requirements of the observations but also information about the progress made towards reaching those requirements.  This information would then be used by the organizers to help determine when to continue the observations and for what duration.

\subsection{Data Transport Management}
Unlike the previous two components to the swarm concept, the design and role of the data transport management system is more loosely defined.  This difference arises because of the different possible methods of correlating the individual array elements.  Specifically, this ranges from fully off-line systems, such as the VLBA, where the correlation is done post observation to fully connected systems like the Karl G. Jansky Very Large Array \citep[VLA;][]{vla} where the correlation is done in real-time.  Thus, how the data are stored and on what time frame they are correlated will impact how the data management is implemented.

The case of connected interferometers with dedicated real-time data transport is the simplest.  For these types of facilities the data management can be implemented as part of the element control system and can consist of a single piece of information about the state of the data transport link.  This state information then becomes one of the requirements for successful observations since there is no way to save the data and perform correlation off-line.  However, one can also imagine a more sophisticated implementation where the data management system also performs basic data checks to make sure that the data not only exists but that it is also valid before it is transmitted.  Similarly, the data management system may be used to implement real-time flagging of interference before correlation.

For interferometers such as the LWA that have off-line elements, intermittent network connectivity between the elements, or network bandwidth that cannot transport all data products in real-time, the data management system must ensure that any observations conducted can be recorded and scheduled to be transferred to a central location for correlation.  Thus, the data management system needs to be able to both monitor the available storage at the element as well as be able to schedule any data transfer operations in such a way that they do not impact the operational state of the element.  Since many of these constraints are coupled with the scheduling of the telescope, the data transport management may be partially or fully implemented as part of the element control software.  The LWA data transport management systems is described in more detail in \S\ref{sec:smartcopy}.

\section{Implementation at the Long Wavelength Array}
\label{sec:implementation}
Following the core concepts presented in \S\ref{sec:core} we have implemented the swarm telescope approach on the stations of the Long Wavelength Array.  The LWA is a low frequency radio dipole array operating between 10 and 88 MHz in the state of New Mexico, USA \citep{LWA, FL}.  Upon completion, the full array will consist of 50 stations distributed across the state with baselines up to 300 km.  It will provide arc second resolution and mJy sensitivities at the observed frequencies.  The array currently consists of two stations: LWA1 which is co-located with the VLA, and LWA-SV which is located on the Sevilleta National Wildlife Refuge approximately 60 km northeast of LWA1.  Each LWA station consists of 256 dual-polarization dipole antennas inside a 100 m by 110 m ellipse elongated north-south.  The signals from each of the antennas are digitized and digitally combined to provide both an ``all-sky" mode that provides raw voltages from each antenna along with a beamformed mode that is suitable for correlation between the stations.  Both stations support between two (LWA-SV) and four (LWA1) independent pointing centers for the beams.

The swarm concept was initially deployed at LWA1 to allow real-time triggering to search for prompt emission from gamma ray bursts and gravitational wave triggers \citep{ligo}, and to participate in multi-wavelength campaigns to observe fast radio bursts \citep{frb}.  Initially, a semi-automatic method was used in which observations for the trigger were automatically generated and emailed to the LWA operator.  These observations then had to be manually scheduled by the operator which lead to scheduling delays that were at best 10 minutes and, in some cases, up to several hours.  This large delay was particularly problematic outside of working hours.  Since these projects were searching for prompt emission, these delays rendered the observations useless.  Thus, a fully automated approach was needed.  The requirements for fully automated triggering are easily implemented inside the swarm concept, which includes monitoring the operational state of the station and being able to schedule the trigger observation automatically.  The two main pieces of software that have been developed for this purpose are the Heuristic Automation for LWA1 ({\tt HAL}) systems and {\tt SmartCopy}.  Using these systems has reduced the time required to get on source to only one to two minutes after the notification is received.

\subsection{The {\tt HAL} Systems}
\label{sec:hal}
\begin{figure}
	\centering
	\includegraphics[width=5in]{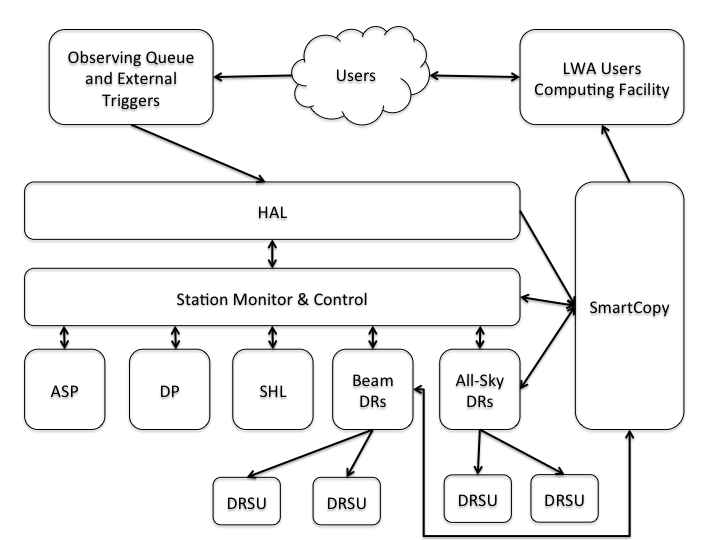}
	\caption{\label{fig:hal}Block diagram showing how {\tt HAL} and {\tt SmartCopy} interact with an LWA station.  The LWA subsystems are: ASP - the analog signal processor, DP - the digital processor, SHL - shelter environmental control, and DRs - the data recorders which write the data to the data recorder storage units (DRSUs) which consist of one or more hard drives.  Users submit observations to the observation queue, which is used by {\tt HAL} to establish the schedule.  {\tt HAL} interacts with the station's monitor and control systems as well as {\tt SmartCopy} to complete observations and monitor the state of the array.  Observations can also be copied to the LWA Users Computing Facility for access by users.}
\end{figure}

The {\tt HAL} systems are the autonomous element control systems that run the LWA stations.  These are designed to function a level above the station's standard monitor and control system and interact with the station software using similar interfaces to human operators.  The relationship between the {\tt HAL} systems and the station is shown in Figure \ref{fig:hal}.  {\tt HAL} is implemented in Python using a multi-threaded framework.  The main thread controls the overall operational state of the stations and relies on several monitoring threads, called sentinels, that monitor individual subsystems that comprise the station.  The sentinels are implemented as basic state machines that use the subsystem sensor values and/or their derivatives to make basic go, no-go determinations about the individual subsystems.  This information is provided to the main thread through a series of callbacks that are triggered during the standard polling cycle.  The main thread then combines these various status updates into the global operational state which is used for determining whether or not the station can be safely operated.  In the event that the station is determined to be in an unsafe condition, e.g., overheating of the electronics shelter, nearby lightning, etc., the {\tt HAL} system will close out any running or scheduled observations and shutdown subsystems as needed.  The {\tt HAL} systems also are designed to allow automated recovery from certain conditions, such as after lightning has not been detected within the vicinity of the station for 30 minutes, in order to further minimize the operational overhead associated with running the stations.  The automatic recovery situations are limited to cases where automated recovery is likely to succeed, for example, weather related conditions.  For all other conditions where it is not always clear what the underlying problem is, e.g., overheating of the shelter when the operational load is low, it was determined that it is safer to not recover than to potentially recover to an unsafe state.  This conservative approach to recovery allows time for the LWA staff to investigate these occurrences and try to identify the underlying problem.

Automated recovery of the station has the potential to impact the safety of people working on site.  In order to address this the {\tt HAL} system is designed to record the state of the subsystems at the time the shutdown condition was triggered.  This state is then used to determine which subsystems to initialize during the recovery.  For example, maintenance on the analog signal processor (ASP) subsystem requires the subsystem to be shutdown and powered off due to the high currents inside the rack.  If a thunderstorm or other shutdown condition occurs during the maintenance then {\tt HAL} will not reinitialize the subsystems when the condition clears to prevent the rack from being energized while it is being worked on.

In addition to the global state, the {\tt HAL} systems also maintain a per-beam state that is used to provide dynamic allocation of resources when processing trigger events.  This is done in order to optimize observing at both stations so they than can be used independently of the LWA interferometer to carry out science that only needs a single station.  This dynamic resource allocation also helps to minimize the impact of triggers on previously scheduled observations by allowing the triggers to be processed using unused resources whenever possible.  It also allows for on-the-fly changes to be made to the system to accommodate the triggers.  One such example is the configuration of the analog signal processor which has multiple filters and gains geared toward different observing frequencies.  For low frequency observations, i.e. below 35 MHz, the station is typically configured to run in a lower ASP gain.  If a trigger is received that should be run in a higher gain mode, the {\tt HAL} systems can adjust the digital gain used for the trigger before it is scheduled in order to compensate for the lower ASP gain.

\subsection{{\tt SmartCopy}}
\label{sec:smartcopy}
The data transport management at the LWA stations is split between the {\tt HAL} systems and a dedicated data transport system called {\tt SmartCopy}.  The LWA stations are designed to record data locally to storage units and then transfer the data off-site for analysis at a later time.  Thus, there are two parts associated with managing the data:  the first is making sure that observations can be recorded and the second is coordinating the copy to off-site storage for processing.  The first part is handled by the {\tt HAL} systems as part of the standard operation of each station and implemented by changing to the data recorder storage unit which has the most free space on each data recorder machine when the station is idle.  The second part of the on-site data management system, the data transport, is handled by {\tt SmartCopy}, a Python-based multi-threaded copy queuing system.  {\tt SmartCopy} is responsible for queuing and executing copies both locally and to remote (off-site) destinations such that they do not interfere with on-going observations.  This is done by monitoring the data recorder machines (DRs) for disk writing operations and by implementing the copies using the {\tt rsync} utility\footnote{\url{https://rsync.samba.org/}} so that the copies can be paused and resumed with minimal overhead.  The queuing is implemented independently for each data recorder machine such that copies can be initiated when any part of the station is idle.  The primary destination for the copies is the LWA Users Computing Facility (UCF), a cluster of six computers, where observers can analyze their data.  After the data has been successfully copied, the original file located at the site is queued for automatic deletion.  This step is done both to reclaim recording space on the DRs as well as to reduce fragmentation on the file systems by deleting the files in batches of at least 1 TB.

{\tt SmartCopy} can also support data transport to remote sites for the purposes of automatic archiving through secure shell tunneling with {\tt rsync}.  For {\tt SmartCopy} operations to destinations that are off-site, the software implements a global network lock such that multiple off-site copies do not compete for the available bandwidth.  The primary motivation for this feature is that future LWA stations could be situated in areas that do not initially have access to high-speed networking infrastructure.

\subsection{Autonomous Operation of the Stations}
\label{sec:lwaops}
\begin{figure}
	\centering
    \includegraphics[width=5in]{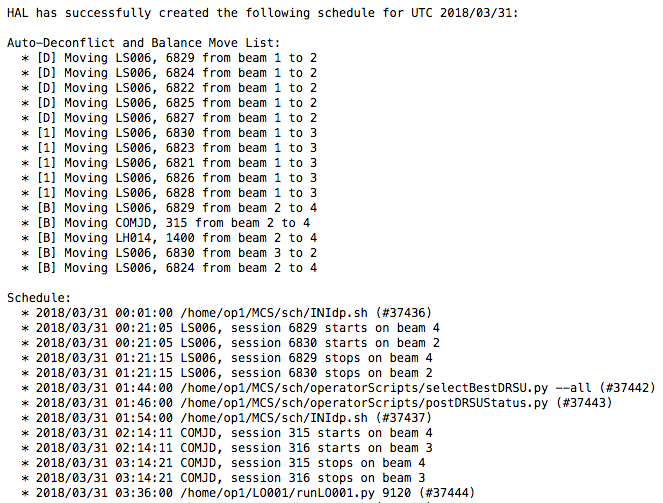}
    \caption{\label{fig:schedule}Example schedule notification email for 2018 March 31 sent by the {\tt HAL} system at LWA1.  The first section, the ``Auto-Deconflict and Balance Move List", shows how the observations in the queue for that day are distributed across the station's beams.  The second part, the ``Schedule", lists all of the observation start and stop times along with the necessary operational support commands.}
\end{figure}

The combination of the {\tt HAL} systems and {\tt SmartCopy}, along with an automated framework for LWA users to submit observations, allows for almost completely autonomous operations of both stations as an interferometer or as independent single-dish radio telescopes.  On a typical day:

\begin{counting}
	\item Users submit observations as session definition files (SDFs) for observing to the LWA SDF validator and indicate which station to use.  The SDFs are checked for validity in real-time via a web-based application\footnote{\url{https://fornax.phys.unm.edu/lwa/validator/index.html}} and all valid schedules are queued for the stations.
	\item The {\tt HAL} systems pull SDFs from the validator queue at the end of each UTC day and create the schedule for the following day.  This includes checking for conflicts between observations, scheduling the observations and implementing supporting commands needed for the observations (filter changes, re-initializations, etc.), and executes any maintenance related commands, see Figure \ref{fig:schedule}.  Observations are also allocated across the various beams in order to try to automatically resolve conflicts between different programs and to spread the recording load across the different data recorders.  In addition to explicitly defined observations any ``filler" projects that can be run during idle times are also scheduled.  These consist of projects that require a larger number of observing hours and are not sensitive to the observation time, e.g., searches for meteor afterglows in beamformed data \citep{meteors}.  After the schedule is set, an e-mail is sent to the LWA staff as notification, but no action is required.
	\item The {\tt HAL} systems monitor the health of the stations throughout the day, post updates of the system status, temperatures, current schedule, and remaining recording capacity on the LWA ``OpScreen" pages \footnote{\url{https://lwalab.phys.unm.edu/OpScreen/os.php}} and prepare the observation metadata for archiving.  The {\tt HAL} systems also schedule copies with {\tt SmartCopy} for any users who have requested their data on the LWA UCF.  In addition, data processed through the real-time DR spectrometer is also automatically copied back to the University of New Mexico to be archived in the LWA data archive\footnote{\url{https://lda10g.alliance.unm.edu/}}.
	\item If the {\tt HAL} systems take an action to protect the station or recover from a recoverable event then an e-mail is sent to the LWA staff to alert them of the current state of the system.  Observers with projects that are interrupted by these events are also automatically notified so that they may resubmit them for observation at a later time.
\end{counting}

\subsection{Triggering}
This standard operating procedure continues until a trigger is received.  Once a trigger is received by the stations:

\begin{counting}
	\item The trigger's location and age are first verified to ensure that the trigger is both recent and that the source is above the horizon at the current time.  Triggers failing either of those criteria are rejected.
	\item After a trigger is validated, the state of the telescope is checked in order to see if the observation can be accommodated and, if so, which beam(s) it may run on.  This second check tries to schedule the trigger without negatively impacting any running projects.  If this cannot be done then lower priority projects are canceled in lieu of the trigger.
	\item Once resources have been allocated, {\tt HAL} builds one or more SDFs to observe the triggers and schedules it with the station's monitor and control software.  After it has been successfully scheduled a notification is sent to the LWA staff, the principal investigator for the triggered project, and any observers whose observations may have been canceled.
\end{counting}

This entire trigger process takes between roughly 30 seconds and two minutes, depending on the source of the trigger and the propagation latency.  Most of this time is dominated by generating the schedules and submitting them to the station's monitor and control system, and by the required downtime between observations needed by the DRs.

In addition to the trigger sources noted previously, the triggering system is also used to operate the LWA stations in conjunction with the new 4-band dipoles on the VLA \citep{mjp} as the expanded LWA (eLWA).  This combines the two LWA stations with $\approx$25 VLA antennas to synthesize baselines up to 90 km when the VLA is in A configuration.  This allows for better calibration, increased sensitivity, and a factor of two improvement in the  resolution for the VLA using the LWA stations.  It also brings imaging capabilities to the LWA by adding in additional baselines to the VLA antennas.

The triggering of the eLWA is initiated from the VLA whenever a suitable project is detected.  These triggers are then propagated to the {\tt HAL} systems and scheduled.  Since eLWA observations only require a single beam, the {\tt HAL} systems are able to assign a beam to the triggers dynamically so as to minimize the impact on LWA observers.  After the observations complete the data is automatically queued for copy over to the LWA UCF using {\tt SmartCopy} where it is later correlated.  Thus, in the course of a day the LWA stations may be used as a two element interferometer, work independently on single-dish science, and jointly observe with the VLA all without involvement of the LWA staff.

\subsection{Operational Savings}
\begin{figure}
	\centering
    \includegraphics[width=5in]{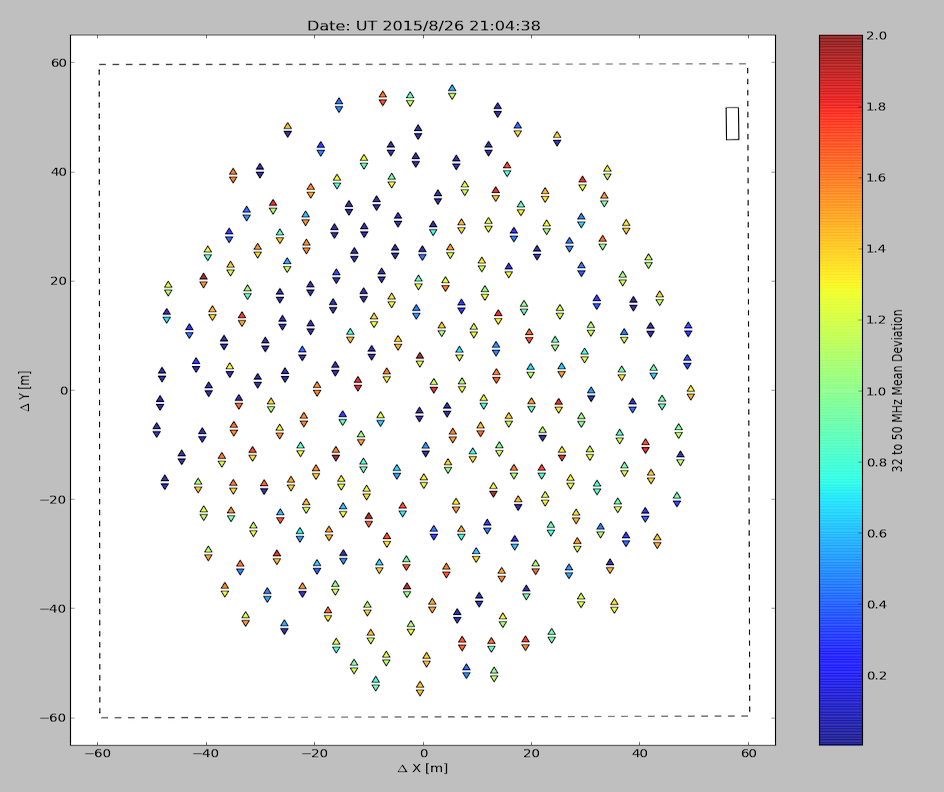}
    \caption{\label{fig:map}Diagram showing the 2018 August 21 lightning damage to the antennas in the LWA1 array.  The figure shows the physical layout of the antennas (triangles and inverted triangles), the perimeter fence (dashed line), and the electronics shelter (rectangle in the upper right corner).  The triangles are color coded by the relative mean power  over 32 to 50 MHz and where blue indicates low power.  The damage can be seen as the blue swath between $(\Delta x,\Delta y)\approx(0,50)$ to $\approx(-50,0)$.}
\end{figure}

The use of the swarm concept has resulted in a significant reduction of operations costs for the LWA.  The primary reason for this is that the {\tt HAL} systems allow for automatic operations of the stations, including data delivery to users, that do not require any manual intervention the vast majority of the time.  This not only reduces direct personnel costs but also ensures that the station will be protected in the event that something occurs, reducing potential damages and the associated maintenance costs.  This happens even if there is a network outage to the station or if the event occurs late at night/early in the morning when human operators are not available.  The best example of this is in the response of the {\tt HAL} system at LWA1 to a thunderstorm that occurred on the afternoon of 2015 August 21.  On this date observations were being made of a pulsar that was being observed through a coronal mass ejection \citep{sun}.  At 22:18 UT lightning was detected in the area and the {\tt HAL} systems halted the pulsar observations and shut down the station.  As part of the lightning shutdown procedure the ASP is powered down which causes the electromechanical relays on all of the input signal paths to physically disconnect.  During this thunderstorm lightning struck the array and damaged approximately 25\% of the front end electronics in the dipoles as shown in Figure \ref{fig:map}.  However, since the ASP was powered off and disconnected there was no direct damage to the ASP or digital processor inside the shelter.  Although there was secondary damage to the ASP caused by trying to power the shorted front ends after the lightning strike, the repairs required were limited to replacing one or two components in the bias-Ts on the ASP boards rather than the entire board.

The swarm approach has also helped reduce maintenance at the stations, both directly through the preventative actions previously presented, as well through the automated data management.  One of the largest sources of maintenance at the stations is replacing failed data recorder storage units (DRSUs) on the DRs which fail at a rate of about one every two to four weeks.  The DRSUs currently consist of five hard drives assembled into a RAID0 array.  Since the data is striped across all five disks the failure of any one can result in data loss.  The {\tt HAL} systems regularly monitor the DRSUs for failing or failed drives and change the active DRSU on the DRs in response to this.  This helps to minimize the loss of data since it is not recorded to known bad locations.  The {\tt SmartCopy} systems further help the situation by moving data off of the DRSUs regularly so that if a DRSU begins to fail the amount of data affected is smaller.  The combination of these two actions reduces the amount of time spent recovering from  these disk drive problems when they fail.  Similarly, the {\tt HAL} systems monitor the outside temperature and automatically shutdown the stations when the temperature becomes too high for the air conditioner units to operate without overheating.  This helps reduce wear on the units and potential damage to the electronics inside the shelter in the event of a catastrophic failure.

\section{Example for the Next Generation Very Large Array}
\label{sec:examples}
The Next Generation Very Large Array, ngVLA, is the future replacement for the VLA being considered by the National Radio Astronomy Observatory.  It is envisioned as a 214 element interferometer of 18m dishes with baselines up to 500 km \citep{ngvla1,ngvla2}.  In addition to the core array, several extensions are being considered.  The two major options are to add longer baselines for very long baseline interferometry (VLBI, the long baseline option) and to add lower frequency receivers along with stand alone dipole arrays to expand the frequency range below 1 GHz (the Next Generation Low Band Observatory, ngLOBO).  \citet{nglobo} have briefly explored how ngLOBO can benefit from being designed with the swarm telescope concept in mind.  In this section we expand the discussion for the ngVLA as a whole with specific examples.

\subsection{Operations and Maintenance}
One of the chief constraints in the design of the ngVLA is trying to minimize the operating costs for the array.  At the same time, the array is being targeted for higher frequency observing, up to 120 GHz, where weather considerations across the several hundred kilometer extent of the array may impact some science programs.  A swarm telescope system with an organizer based communication system would be able to address both of these issues.  First, by moving monitoring and control of each antenna from a centralized human operator to the antennas themselves, the amount of personnel needed for running the array can be reduced.  The sensors and monitoring points necessary for the autonomous element control will also provide detailed information on the status of each dish.  This information can be used either individually or as a whole by machine learning techniques in order to predict failures.  By predicting failures before they render a dish inoperable, preventative maintenance may be better targeted and could also reduce the costs associated with failures.  Second, using an organizer to help create per-project arrays would allow antennas with poor observing conditions or maintenance issues to be automatically excluded.  This reduction in sensitivity could be accounted for in real-time by adjusting the observing time.  Furthermore, the predictive failure information can also be used when scheduling the array so these potential failures can be accounted for.  For example, if the expected number of element failures for a given observation is two, then two additional antennas can be allocated for the observation in anticipation of failure.  If no failures occur then the observation can be stopped once the requested level of sensitivity is met.  Finally, we also note that any remaining unused operational antennas may be used for projects working at lower frequencies where pristine observing conditions are less critical.

\subsection{Scheduling Flexibility}
\label{sec:ngflex}
Beyond simply reacting to observing conditions and the state of the array, the swarm concept can be used to further increase the flexibility of scheduling on the array.  This will not only help with responding to targets of opportunity but will also allow multiple projects to run on the array at the same time.  For this added flexibility to be fully realized there must be a shift in how the array is scheduled.  This shift is from a fixed number of hours to a set of configuration requirements based on the scientific goals of the project.  The configuration requirements would include such things as the minimum number of elements for the desired $(u,v)$ coverage, sensitivity and $(u,v)$ range sampled.  This more fine grained approach would then allow the array to be divided into smaller sub-arrays as necessary.  For example, one could imagine a hypothetical situation where three projects could be run simultaneously.  A VLBI observation could use the longest baselines, while a pulsar study uses the core, and a visibility based transient search, such as a system like Realfast \cite{realfast}, uses the intermediate baselines.  Furthermore, this increased flexibility in how the ngVLA could be scheduled will allow LWA-style ``filler'' projects to be scheduled in addition to commensal observations.

\subsection{Advanced Observing Modes}
In addition to the potential advantages for operations, maintenance, and scheduling, the swarm concept also allows the possibility of advanced observing modes. For example, an organizer based system may be beneficial for the long baseline option by providing a way for groups of four to six 18m dishes to be phased up to provide the sensitivity of a single, larger dish. These groups could provide the sensitivity required for the ngVLA to observe faint supermassive binary black hole systems and to characterize their orbits \cite{bansal17} leading to an improved understanding of the evolution of black holes and galaxies.  We note that grouping the antennas also improves the baseline sensitivity which can improve the calibration achieved and allow the use of fainter and therefore more nearby calibrators.  This grouping can be done in an automated fashion using a second set of dedicated organizers that work with small clusters of geographically close antennas.  These organizers would then initiate and maintain the phasing of the antennas in the cluster based on a combination of the observation goals and the operational status of antennas within the group.  Creating phased clusters of antennas also has some advantages for the data transport and correlator requirements since it can be an effective way of reducing the incoming bandwidth for correlation.

Another possibility is to use the swarm telescope to allow self-triggering, specifically by using the ngLOBO-low dipole array stations which would be able to image the entire sky at low resolution.  In this case several of the dipole arrays would run a fast transient detection over the sky using a fast direct imaging approach, such as the E-field Parallel Imaging Correlator \citep{epic}.  Triggers from each station would be combined by the organizer for anti-coincidence to rule out local RFI and valid triggers would be sent to the full ngLOBO or ngVLA to be triggered for follow-up.  For the ngLOBO-low stations the trigger could be observed using one of the multiple beams available.  The determination to trigger the full ngVLA would be made by the organizer based on the state of the array and which projects were currently observing.

\subsection{Resource Optimizations}
The dynamic resource allocation that is afforded by a self-organizing swarm telescope may lead to new ways to optimize observations for operations-based considerations such as power consumption, maintenance on antennas, input bandwidth, and correlator resources.  For example, the resources utilization for an FX correlator with $N$ elements is $O(N^2)$.  By running three projects, as in the case of the scheduling example discussed in \S\ref{sec:ngflex}, each project needs only $O(N^2/9)$ and the correlator would need roughly one-third of the resources compared to correlating the full array.  This savings could translate to a reduction in the power consumption of the correlator.  Similarly, by using the phased clusters the value of $N$ can be further reduced while still maintaining the sensitivity of the full array.  Furthermore, such an implementation will reduce data transport requirements of the array.

\section{Conclusion}
\label{sec:conclusion}
We have presented an operating concept for radio interferometers called the swarm telescope concept based on its resemblance to the idea of swarm intelligence where by multiple independent systems come together to give rise to more advanced behavior.  We have also provided a concrete example of an implementation of the swarm telescope concept as it is used on the stations of the LWA.  Furthermore we have explored the specific case of how the ngVLA can take advantage of these concepts and potentially use them to reduce operations costs for the array and improve responsiveness to targets of opportunity.

In general, we see the swarm concept as a natural extension to advances in robotic and thinking telescopes, combined with interferometry.  It allows for more efficient operations of facilities by moving much of the daily operational work done by humans to autonomous control systems.  This, in turn, frees up personnel to focus on the scientific output of the telescope.  The swarm concept can also combine the unused resources of the different elements together to form an {\textit ad hoc} array.  This concept is particularly powerful for low frequency dipole arrays where the ability to form multiple beams allows for both single station science as well as interferometric observations to be conducted simultaneously.  This feature offers a unique approach for funding and constructing small, collaborative interferometers by allowing different elements to be owned and operated by different entities.  Under this concept each member of the collaboration benefits from being able to not only run their own single-dish science while also being able to perform interferometric observations.

Further developments of this approach may also be useful out of the radio interferometer context and be applied to facilities at other wavelengths.  In particular the area of dynamic resource allocation may be of interest to the optical community since an entity may operate many telescopes on the same mountain top.  This would allow queue observing to not only optimize for observing conditions but also to match the goals with the available instrumentation across many telescopes.  This may also be applicable for collaborative sharing of time between different telescopes.

\section*{Acknowledgements}
We thank the anonymous referee for many helpful suggestions for improving the manuscript.  Construction of the LWA has been supported by the Office of Naval Research under Contract N00014-07-C-0147. Support for operations and continuing development of the LWA1 is provided by the National Science Foundation under grant AST-1139974 of the University Radio Observatory program.

\bibliographystyle{ws-jai}
\bibliography{ms} 

\clearpage

\clearpage

\clearpage

\clearpage

\end{document}